\documentclass[runningheads]{llncs}
\usepackage{booktabs}
\usepackage{graphicx}
\usepackage{adjustbox}
\usepackage{multirow}
\usepackage[table,xcdraw]{xcolor}
\usepackage{url}

\usepackage{cite}
\usepackage{graphicx} 
\usepackage{subfigure}
\usepackage{amsmath}
\usepackage[noend]{algpseudocode}
\usepackage{algorithmicx,algorithm}
\usepackage{float}
\usepackage{booktabs}
\usepackage{multirow}
\usepackage{hyperref}
\usepackage{xcolor}

\usepackage{xcolor}
\newcommand{\hyss}[1]{\textcolor{black}{#1}}

\newcommand{\hys}[1]{\textcolor{black}{#1}}
\newcommand{\wqz}[1]{\textcolor{black}{#1}}

\newcommand{\eg}{\textit{e.g.}}

\newcommand{\ie}{\textit{i.e.}}
\begin{document}
%
\title{Rethinking Degradation: Radiograph Super-Resolution via AID-SRGAN
} 
%
%
\author{Yongsong Huang\inst{1} Qingzhong Wang~\inst{2} \and~Shinichiro Omachi\inst{1}}
%
\authorrunning{Yongsong Huang et al.}
%
\institute{Tohoku University, Sendai, Japan\\ \email{huang.yongsong.r1@dc.tohoku.ac.jp,	machi@ecei.tohoku.ac.jp} \and Baidu Research, Beijing, China.\\ \email{wangqingzhong@baidu.com} }
%
\maketitle              
\begin{abstract}


In this paper, we present a medical \textbf{A}ttent\textbf{I}on \textbf{D}enoising \textbf{S}uper \textbf{R}esolution \textbf{G}enerative \textbf{A}dversarial \textbf{N}etwork (AID-SRGAN) for diographic image super-resolution.
First, we present a 
medical practical degradation model that considers various degradation factors beyond downsampling. To the best of our knowledge, this is the first composite degradation model proposed for radiographic images.
Furthermore, we propose AID-SRGAN, which can simultaneously 
denoise 
and generate high-resolution (HR) radiographs. 
\hyss{In this model, we introduce an attention mechanism into the denoising module to make it more robust to complicated degradation.}
\wqz{Finally, the SR module reconstructs the HR radiographs using the ``clean'' low-resolution (LR) radiographs. In addition, we propose a separate-joint training approach to train the model, and extensive experiments are conducted to show that the proposed method is superior to its counterparts. \eg, our proposed method achieves 31.90 of PSNR with a scale factor of $4\times$, which is 7.05\% higher than that obtained by recent work, SPSR ~\cite{ma2021structure}. Our dataset and code will be made available at: \href{https://github.com/yongsongH/AIDSRGAN-MICCAI2022}{https://github.com/yongsongH/AIDSRGAN-MICCAI2022}.}
\end{abstract}

\section{Introduction\label{sec1}}

High-resolution musculoskeletal radiographs provide more details that are crucial for medical diagnosis, particularly for diagnosing primary bone tumors and bone stress injuries~\cite{musculoskeletal2020,von2021multitask,bone2021,Moran2021SR,miccaiPengZC21}. However, radiographic image quality is affected by many factors, \wqz{such as scanning time, patients' poses, and motions}, and achieving higher-resolution medical images is expensive and time-consuming because it requires a relatively long scanning time. 
\hyss{However, existing SR algorithms fail to fully consider the degradation factors mentioned above.}

\hyss{Imperfect degenerate models put the algorithm at risk of domain shift (see Table.\ref{tab.2}: (a) Domain shift). To solve this problem,} we first need to \hyss{rethink} the degradation -- high quality transforms into lower quality. The degradation of radiographs is related to statistical noise, external disruptions, and downsampling~\cite{CT2019review,Mohan2020SABERAS,wang2021real,motion2020patient}. First, the two basic types of 
statistical noise -- Poisson and Gaussian–are common in radiographic images~\cite{CT2019review,GuanLCWLZ21}. However, most of the \wqz{existing} 
state-of-the-art deep learning-based SR methods focus only on the degradation of bicubic downsampling ~\cite{Rezaei2017DeepLF,ma2021structure,huang2021infrared}, \wqz{\ie, the models directly take the bicubic downsampling images as input and reconstruct the HR images, leading to the problem of domain shift when applied to noisy images}. Second, the disruptions in the \hyss{application scenarios} arise from the following factors: radiologists and patients, as well as information loss due to compressed transmission 
via the Internet ~\cite{PengYLY20,Dimililer2022DCTbasedMI}. \hyss{In general,} operational mistakes and the patients' displacement relative to the device would introduce motion blur~\cite{Rezaei2017DeepLF,honarvar2021motion}. On the other hand, telemedicine~\cite{Heller2021EducationalCA} requires uploading medical images to a data center for easy consultation and online storage. Owing to the limited network bandwidth, 
compression~\cite{Dimililer2022DCTbasedMI} is widely employed for online data transmission, 
leading to low-quality radiographic images. 


To address this problem, more attention has been paid to the tasks of medical image super-resolution\cite{miccaiPengZC21,de2021impact,van2020super}. 
Deep learning-based methods~\cite{Zhang2018LearningAS,Zhang2019DeepPS,huang2021infrared,Wang2018ESRGANES,ma2021structure} 
\wqz{dominate image SR, which learns a mapping} from LR images to HR images and differs from 
traditional methods in which more prior knowledge is required~\cite{Chen2022RealWorldSI}. 
\wqz{Recently, blind SR that considers real-world degradation has drawn much attention~\cite{Zhang2018LearningAS,Zhang2019DeepPS,wang2021real,Wang2021UnsupervisedDR} because it is more common. However, only Gaussian blur and compression were adopted to synthesize the paired training data \cite{wang2021real} and the parameters of the Gaussian kernels were fixed~\cite{Zhang2018LearningAS,Zhang2019DeepPS}. In fact, the degradation of radiographs could be more complicated, including statistical noise, motion blur, and compression.} In this study, we aim to address the problem of musculoskeletal radiograph SR. \wqz{The main contributions are in three-fold.}
\begin{itemize}
    \item We propose a 
    practical degradation model for radiographs, \wqz{which considers most possible degradation factors, such as statistical noise, motion blur, compression, and each of them has variant parameters. \hyss{ This model aims to represent complex nonlinear degeneracies, unlike current models that focus more on downsampling.} In addition, the degradation model is applied to synthesize data to train the proposed SR model.}
    \item \wqz{We propose a medical attention denoising SRGAN model (AID-SRGAN). An attention mechanism is introduced into the denoising module to make it more robust to complicated degradation. Moreover, we propose a two-stage training approach to train the proposed AID-SRGAN, \ie, we first separately train the denoising module and SR module to obtain a relatively good denoising network and SR network, respectively. We then jointly train the denoising and SR modules in an end-to-end manner to further improve the performance. \hyss{Finally, it is a flexible framework and easy to follow.}}
    \item \wqz{We conduct extensive experiments and compare the proposed model with \hys{other existing works}. AID-SRGAN is superior to its counterparts, achieving 31.90 dB of PSNR with a scale factor of $4\times$, which is 7.05\% higher than a recent work -- SPSR~\cite{ma2021structure}. Moreover, considering SSIM, AID-SRGAN outperforms SPSR, \eg, 0.9476 vs. 0.9415. In addition, ablation studies on the modules and hyper-parameters are conducted to demonstrate their effectiveness.}
\end{itemize}


\section{Methodology\label{sec2}}


\begin{figure}[t]
\centering
\includegraphics[width=0.75\textwidth]{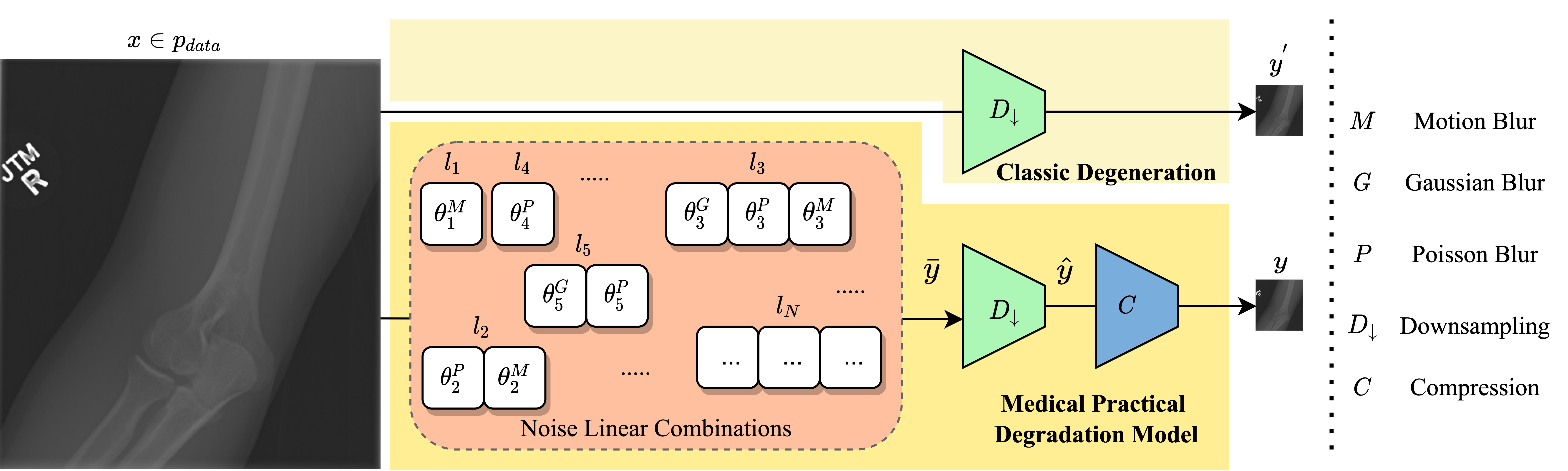}
\caption{Overview of the medical practical degradation model. 
\wqz{The model takes $x$ -- HR radiographs as input and imposes $L_{N}$ -- noise linear combinations on $x$, then bicubic downsampling and compression are applied to the noisy radiographs, yielding noisy LR radiographs $y$. We also generate LR radiographs $y^\prime$ via directly imposing downsampling on HR radiographs.}
}
\label{fig1}
\end{figure}

\subsection{Medical Practical Degradation Model\label{sec2.1}}

In the real world, HR images $x$ are degraded to LR images $y$, with random and complicated degradation mechanisms~\cite{wang2021real,Wang2021UnsupervisedDR}. An HR image suffers from blur kernels $k$, such as statistical noise, which first degrades the image quality. Furthermore, the damaged image $\bar{y}$ is transformed into an LR image $\hat{y}$ by downsampling $D_{\downarrow}$. Finally, image compression $C$ is \wqz{used for online transmission and storage. The entire procedure is represented using the following function:}

\begin{equation}
y=\mathcal{F}^{D}(x)=C\left(D_{\downarrow}(x\oplus  k)\right),
\end{equation}
where $\mathcal{F}^{D}$ denotes a degradation function. However, existing SR models only consider one or two degradation factors~\cite{ma2021structure,Wang2018ESRGANES,Zhang2019DeepPS,huang2021infrared}, 
\wqz{which could result in a large gap between \hyss{open world} LR images and synthetic LR images~\cite{wang2021real} and lead to poor performance in practice due to the domain-shift problem.}

In this study, we propose a practical degradation model for radiographs ( Fig. ~\ref{fig1}), 
\wqz{where we consider most degradation factors in radiographs -- statistical noise combinations $L_{N}$, downsampling $D_{\downarrow}$, and compression $C$. In $L_N$, we combine the Gaussian blur $G$, Poisson blur $P$ and motion blur $M$, which are widely observed in radiographs.}
In particular, the statistical noise parameters $\theta^{L_{N}}$, such as expectation $\mathbf{\mu}$ and variance $\mathbf{\Sigma}$, are automatically updated when a new sample $x$ is fed into the degradation model, \wqz{yielding a noisy image $\bar{y}$, \ie,
\begin{equation}
    \bar{y} = L_N\left(x; \theta^{L_n}\right)\quad \theta^{L_N}\sim p\left(\theta^{L_N}\right).
\end{equation}
Then, $\bar{y}$ is resized using downsampling and compressed to generate noisy LR images $y$.}




\wqz{From the perspective of latent space~\cite{Lee2022LearningMP}, the proposed medical practical degradation model 
applies external variables $\theta$ and latent variables $\bar{y}, \hat{y}$ 
to approximate the real-world degradation distribution $p(y|x)$, \ie, $\tilde{p}(y \mid x) \approx p(y \mid x)$, where $y$ and $x$ are LR and HR images, respectively. 
The distribution can be computed as follows:
\begin{equation}
    \tilde{p}(y\mid x)=\int p(y\mid \hat{y},\theta)p(\hat{y}\mid \bar{y},\theta)p(\bar{y}\mid x,\theta)p(\theta)d\hat{y}d\bar{y}d\theta.
\end{equation}
}

\wqz{We draw samples $y$ from $\tilde{p}(y|x)$ and build LR-HR pairs $\{(y_1, x_1), \cdots, (y_n, x_n)\}$ to train AID-SRGAN. \hyss{This degradation model is expected to attract researchers to focus on medical image sample representation more. }}



\begin{figure}[t]
\centering
\includegraphics[width=0.8\textwidth]{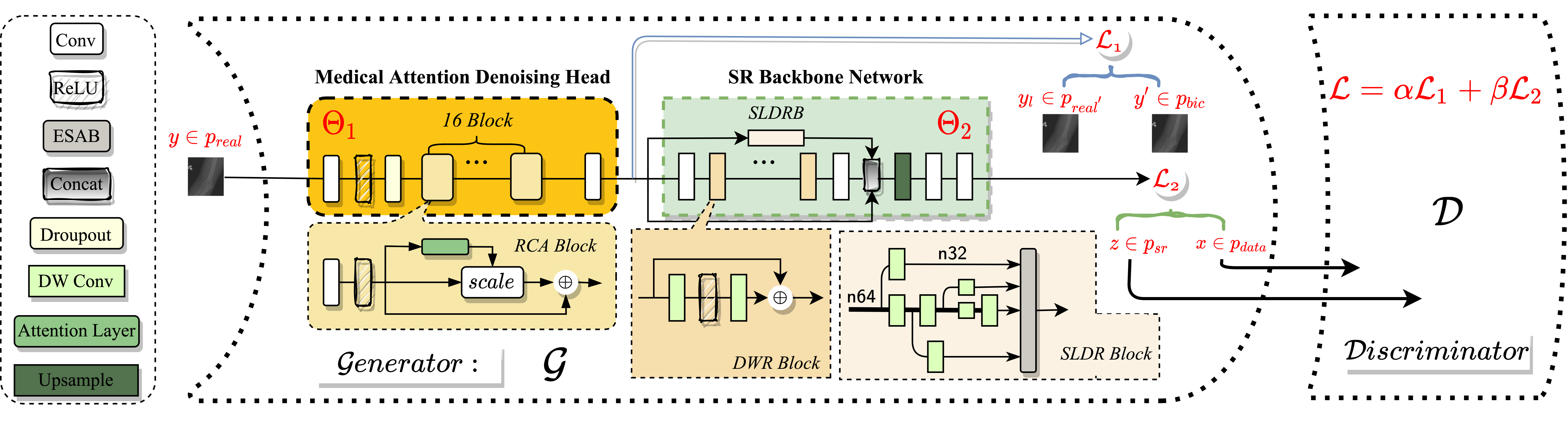}
\caption{Overview of our proposed AID-SRGAN. We can jointly train medical attention denoising head and SR backbone network in an end-to-end manner. AID-SRGAN with the input $y \in p_{real}$, HR image $x \in p_{data}$ and $z \in p_{sr}$ -- the output fake sample by generator $\mathcal{G}$. $\mathcal{D}$ is the discriminator. $y^\prime\in p_{bic}$ denotes the sample obtained using downsampling degradation only. $\mathcal{L}$ is a loss function. $\Theta_{1}$ and $\Theta_{2}$ represent the parameters of $\mathcal{G}$. \hys{$\alpha$ and $\beta$ are hyperparameters.} Zoom in for best view.}
\label{fig3}
\end{figure}

\subsection{Medical Attention Denoising SRGAN (AID-SRGAN)\label{sec2.2}}

\hyss{For AID-SRGAN, our goal is to propose a straightforward model that is easy to follow.} After obtaining the training data, 
there are \hyss{different} approaches to 
solve the real-world super-resolution problem, \hyss{such as 1) denoising first and then reconstructing HR images, or 2) direct reconstruction ~\cite{Zhang2018LearningAS,Zhang2019DeepPS,wang2021real,Wang2021UnsupervisedDR}.} 
\wqz{The proposed AID-SRGAN adopts denoising first and then reconstructs the HR images. \hys{Moreover, denoising first sounds more reasonable because LR images are noisy.} We demonstrate the framework of the AID-SRGAN in Fig. ~\ref{fig3}, which is composed of two modules: denoising, which reconstructs the LR images obtained by downsampling only, and super-resolution, which reconstructs the HR image from the reconstructed ``clean'' LR image.} \hyss{In summary, the denoising module takes noisy LR images $y$ as input and reconstructs $y^\prime$, while the SR module takes $y^\prime$ as input and reconstructs HR images $x$. \hyss{For training strategies, we can separately train the two modules to obtain a relatively good initialization of AID-SRGAN and then train the two modules in an end-to-end manner via backpropagation.}}

\begin{figure}[t]
\centering
\includegraphics[width=0.75\textwidth]{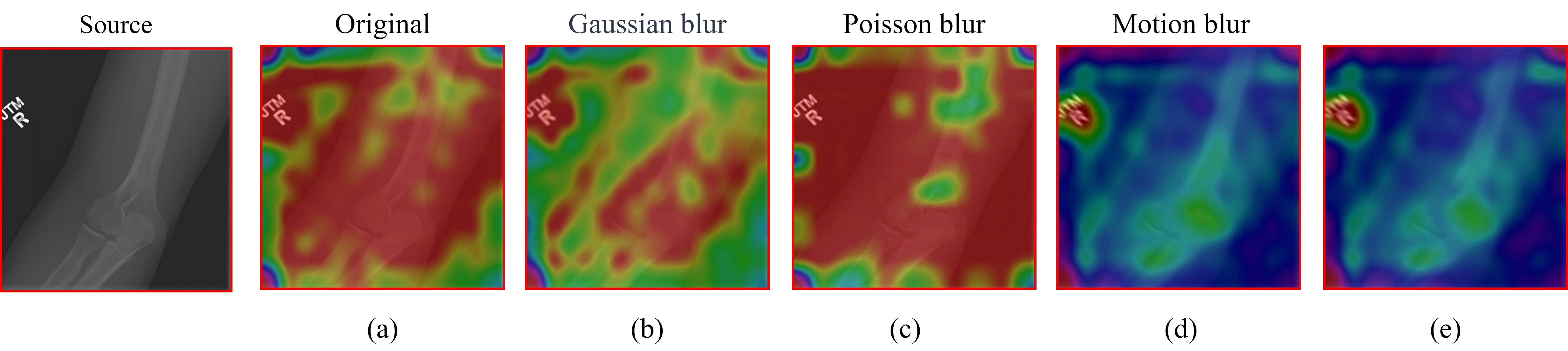}
\caption{Visualization of the heat map for the degradation representation with different factors by CAM~\cite{Selvaraju2019GradCAMVE}. (e) illustrates representations generated by motion blur, Gaussian blur, and compression. \hyss{It is necessary to guide the denoising, considering the model's different response to various degradation factors.} Best viewed in color.}
\label{fig2}
\end{figure}

\textbf{Medical attention denoising:} \wqz{one observation of the denoising deep neural networks is that the activation map varies using different degradation factors. Fig.~\ref{fig2} presents the activation map with different degradation kernels, and it can be observed that the network pays more attention to objects, such as characters using motion blur, while focusing on the entire image using Gaussian blur. \hyss{In summary, there will be different responses for different degradation factors.} To obtain consistent representations and adapt to different degradation factors, we introduce an attention mechanism~\cite{Hu2020SE} to guide the denoising procedure.}
We adopted the residual channel attention (RCA)block as the basic unit for the attention denoising head, 
\wqz{which is calculated as follows:
\begin{equation}
    x^{l+1} = x^l + \mathbf{\sigma}\left(Conv(x^l)\right)\odot x^l,
\end{equation}
where $x^l$ denotes the $l$th layer input, $Conv(\cdot)$ represents the convolution layer, $\mathbf{\sigma}(\cdot)$ denotes the activation function, and $\odot$ represents the element-wise multiplication.}
In contrast to other methods, our network does not estimate the blur kernel, which is beneficial for reducing reliance on extensive prior knowledge~\cite{Son2021TowardRS}.

\textbf{SR  backbone network:} 
When the distribution of $y_{l}$ is available, we seek to reconstruct $z \in p_{sr}$ from $y_{l} \in p_{real^{'}}$ using an adversarial training approach. We also observed that PSRGAN\cite{huang2021infrared} performs reliably in image reconstruction for images with simple patterns and structures, such as infrared and gray images, which benefits radiological image reconstruction. 

\begin{equation}
\begin{aligned}
&\min _{\mathcal{G}} \max _{\mathcal{D}} V(\mathcal{D}, \mathcal{G})=E_{x \sim p_{\text {data }}}[\log \mathcal{D}(x)] \\
&+E_{y_{l}  \sim  p_{\text {real}^{'}}, y \sim p_{\text {real }}}\left[\log \left(1-\mathcal{D}\left(z=\mathcal{G}\left(y_{l} \mid y\right)\right)\right)\right].
\end{aligned}
\label{eq.6}
\end{equation}

\hyss{For the generator $\mathcal{G}$, the output (from the denoising head) is fed to the SR backbone network. The backbone network consists of the main and branch paths, which are built using DWRB and SLDRB, respectively.
The detailed information is shown in Fig.\ref{fig3}, where the key component-SLDRB in the branch is introduced the information distillation. This distillation method improves the feature representation of the model by setting different numbers of feature channels (as shown in Fig.\ref{fig3}: n64 \& n32), which is believed to be beneficial for images with fewer patterns in the experiments\cite{huang2021infrared}.} The discriminator $\mathcal{D}$ is trained to maximize the probability of providing a real sample to both training data and fake samples generated from $\mathcal{G}$. $\mathcal{G}$ is trained to minimize $\log \left(1-\mathcal{D}\left(z=\mathcal{G}\left(y_{l} \mid y\right)\right)\right)$, \hyss{where $y_{l} \in p_{real^{'}}$ denotes the bicubic LR input}. The objective function is defined by Eq.\ref{eq.6},  where $V$ denotes the divergence. $E$ represents cross-entropy.

\subsection{Training Strategy \label{sec2.3}}

\wqz{We first separately train the denoising network and SR network using the synthetic data, \ie, the paired data $(y, y^\prime)$ is employed to train the denoising network, and $(y^\prime, x)$ is used to train the SR network. In summary, we seek to determine the optimal parameters $\theta^{*}$ by minimizing the expected risk. }

\begin{equation}
\theta^{*}=\underset{\theta \in \Theta_{1}}{\operatorname{argmin}} \mathbf{E}_{y, y^\prime}\left[\mathcal{L}_{1}\left(y, y^\prime, \theta )\right]\right.
\end{equation}

where $\mathcal{L}_{1}\left(y, y^\prime, \theta \right)$ is a loss function that depends on parameter $\theta$. $y$ and $y^\prime$ denote the input image and bicubic downsampling image, respectively.
After separate training, we employ $(y, x)$ to jointly train the entire framework in an end-to-end manner. Separate training obtains a good local optimal of AID-SRGAN, and joint training further boosts the performance, which is similar to the pre-training-fine-tuning paradigm~\cite{devlin2018bert}; however, we employ supervised pre-training. \hyss{According to the pre-experimental results, the medical attention denoising head performs better when compared with the direct use of DNCNN denoising or DNN-based denoising heads (see Table \ref{tab.2}: (b) Network selection \& (c) RCA Block).}


\begin{table}[t]
\centering
\caption{PSNR$\uparrow $ and SSIM$\uparrow $ results of different methods on MURA-mini (mini)\& MURA-plus (plus)  with scale factors of  4 \& 2. \hyss{Ours and Ours+ have 16 and 256 RCA blocks, respectively. There are two test datasets with the same HR images but different degraded LR images (which are more damaged).} The best results are in \textbf{bold}.}
\label{tab2}
\resizebox{0.8\textwidth}{!}{%
\begin{tabular}{@{}ccc|cc|cccccc@{}}
\toprule
                                                 & \multicolumn{2}{c|}{}                                  & Ours+           & Ours          & Bic    & DPSR & PSRGAN & SRMD   & ESRGAN & SPSR   \\ \midrule
\multicolumn{1}{c|}{\multirow{4}{*}{$\times 4$}} & \multicolumn{1}{c|}{\multirow{2}{*}{mini}} & PSNR & \textbf{31.90} & 31.21           & 28.55  & 29.79  & 29.39  & 16.32  & 28.63  & 29.80  \\ \cmidrule(lr){3-3}
\multicolumn{1}{c|}{}                            & \multicolumn{1}{c|}{}                           & SSIM & 0.9476         & \textbf{0.9506} & 0.9403 & 0.9497 & 0.9306 & 0.6247 & 0.9313 & 0.9415 \\ \cmidrule(l){2-11} 
\multicolumn{1}{c|}{}                            & \multicolumn{1}{c|}{\multirow{2}{*}{plus}} & PSNR & \textbf{31.52} & 30.78           & 28.68  & 29.98  & 28.65  & 16.31  & 28.07  & 28.77  \\ \cmidrule(lr){3-3}
\multicolumn{1}{c|}{}                            & \multicolumn{1}{c|}{}                           & SSIM & 0.9454         & \textbf{0.9469} & 0.9354 & 0.9431 & 0.9273 & 0.6204 & 0.9267 & 0.9358 \\ \midrule
\multicolumn{1}{c|}{\multirow{4}{*}{$\times 2$}} & \multicolumn{1}{c|}{\multirow{2}{*}{mini}} & PSNR & 34.03          &\textbf{34.11}           & 29.91  & 31.00  & 29.87  & 14.75  & 30.03  & 30.74  \\ \cmidrule(lr){3-3}
\multicolumn{1}{c|}{}                            & \multicolumn{1}{c|}{}                           & SSIM & 0.9589         & \textbf{0.9590}          & 0.9440 & 0.9412 & 0.9470  & 0.5571  & 0.9438 & 0.9357  \\ \cmidrule(l){2-11} 
\multicolumn{1}{c|}{}                            & \multicolumn{1}{c|}{\multirow{2}{*}{plus}} & PSNR & \textbf{32.54}         & 32.43           & 30.21  & 31.59  & 29.04  & 14.74  & 29.43  & 30.53 \\ \cmidrule(lr){3-3}
\multicolumn{1}{c|}{}                            & \multicolumn{1}{c|}{}                           & SSIM & 0.9473         &\textbf{0.9483}         & 0.9361 & 0.9322 & 0.9381  & 0.5570  & 0.9340 & 0.9313 \\ \bottomrule
\end{tabular}%
}
\end{table}

\section{Experiments\label{sec3}}

\textbf{Dataset:}\label{sec3.1}
we employed a widely used dataset, MURA~\cite{rajpurkar2017mura} to synthesize training pairs. MURA contains 40,005 musculoskeletal radiographs of the upper extremities. 
\wqz{We selected 4,000 images as the training set, named MURA-SR. Two test datasets were used, MURA-mini and MURA-plus. Both were composed of 100 HR images and different degraded LR images. For MURA-SR, the blur kernel size was randomly selected from $\left \{ 1,3,...,11 \right \}$, whereas MURA-mini and MURA-plus used kernel sizes selected from $\left\{ 1,3,5 \right \}$ and  $\left\{7,9,11 \right \}$ ,respectively, \ie, the degradation of MURA-plus is more serious. The probability of using a blur kernel was randomly selected from $\left \{ 0.1,0.2,...,1.0 \right \}$. Finally, the JPEG compression quality factor was set to 3~\cite{liu2018non,wang2021real} for all images.}

\begin{figure}[t]
\centering
\includegraphics[width=\textwidth]{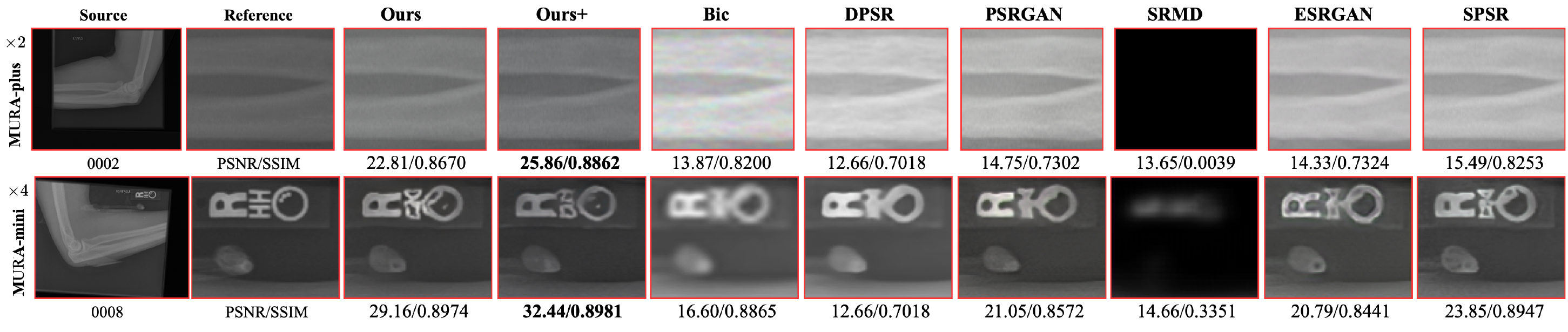}
\caption{\textbf{Top:} Qualitative comparisons on 002 samples from MURA-plus with upsampling scale factor of 2. \textbf{Bottom:} Qualitative comparisons on 008 samples from MURA-mini with upsampling scale factor of 4. The PSNR$\uparrow$/SSIM$\uparrow$ of the test images are shown in the figure. The best results are in \textbf{bold}. Zoom in for best view.}
\label{fig4}
\end{figure}

\textbf{Training details:} We trained our model with a batch size of 32 on two TITAN X (Pascal) GPUs. The training HR patch size was set to 96. We employed the Adam optimizer~\cite{kingma2014adam} with a learning rate of $1e- 5$. We used VGG16~\cite{simonyan2014very} as the discriminator and the combination of L1 and SSIM as the loss function~\cite{wang2021real}. 

\subsection{Results\label{sec3.4}}

\textbf{Quantitative results:} in experiments, we propose two versions of AID-SRGAN: \textit{Ours} (16 RCA Blocks) and \textit{Ours+} (256 RCA Blocks). 
As shown in Table \ref{tab2}, we compare our approach with downsampling-based models~\footnote{The degradation model only consider downsapling.} (PSRGAN~\cite{huang2021infrared}, ESRGAN~\cite{Wang2018ESRGANES},SPSR~\cite{ma2021structure}), and real-world oriented models~\footnote{The degradation model considers downsampling and others, such as Gaussian blur.} (DPSR~\cite{Zhang2019DeepPS}, SRMD~\cite{Zhang2018LearningAS}), and \wqz{we trained all the models using MURA-SR}.
The metric scores (PSNR and SSIM) were calculated for the Y channel of the YCbCr space. 
\wqz{We can easily conclude that our proposed model is superior to the counterparts on both MURA-mini and MURA-plus with the upsampling factors of 4 and 2 considering PSNR and SSIM, \eg, Ours+ achieves 31.90 dB of PSNR on MURA-mini with the scale factor of 4, which is roughly 7.08\% higher than that achieved by DPSR. Compared with downsampling-based methods (i.e., PSRGAN and ESRGAN)our proposed approach achieves a more remarkable performance, and the average relative improvement is approximately 10\%, indicating that denoising matters in radiograph super-resolution. When comparing Ours+ and Ours, it is evident that deeper denoising modules achieve a higher PSNR, whereas shallow denoising networks obtain a slightly higher SSIM. A possible reason is that shallow networks converge faster than deep networks, and we use SSIM in the loss function; hence, shallow networks can reach a higher SSIM score with the same training epochs.}

\textbf{Qualitative results:} \wqz{Fig. \ref{fig4} presents some examples of the reconstructed HR radiographs generated by different models. Compared with the existing models, the proposed AID-SRGAN can generate a sharper HR image and reconstruct more details, \eg, the edge of the bone in the first image, and the characters in the second image, and the PSNR and SSIM scores are also relatively high compared to the counterparts.}


\subsection{Ablation Studies\label{sec3.3}}

We conducted extensive ablation studies on different modules and hyperparameters in the AID-SRGAN, and the comparison is presented in Table~\ref{tab.2}. \hyss{In complex degenerate models, the SR algorithm will have difficulties fitting the data distribution. This also explains the domain drift (see (a), domain shift), and paying more attention to the degradation model is beneficial.} In (b), network selection, we can see that SR+ denoising directly outperforms SR using DNCNN\cite{Zhang2017dncnn} as the denoising network. The performance can be boosted to 28.78 dB using the RCA block, as shown in (c), while the ablation experiments with hyperparameters are shown in (d). Using dropout, we can further improve the performance, achieving 29.34 dB of PSNR. Finally, we jointly trained the denoising and SR networks based on the separate pre-training models, achieving 30.14 of PSNR and 0.9271 of SSIM.

\begin{table}[t]
\caption{Ablation on using more solutions (network structure and test dataset).   SR network is PSRGAN. The denoising model is DNCNN. \hyss{More setting details as follows: upsampling factor: $\times 4$, evaluated in RGB space.}}
\label{tab.2}
\begin{minipage}[b]{.31\linewidth}
\resizebox{\textwidth}{!}{%
\centering
\begin{tabular}{@{}ccc@{}}
\multicolumn{3}{c}{\textbf{(a) Domain shift.}}                             \\ \midrule
\multicolumn{1}{c|}{\begin{tabular}[c]{@{}c@{}}Test \\ dataset\end{tabular}} & \multicolumn{1}{c|}{\textbf{PSNR/$dB $}} & \textbf{SSIM$\uparrow $} \\ \midrule
Bicubic   & 33.66                         & 0.9383                         \\
MURA-mini & \cellcolor[HTML]{C0C0C0}26.85 & \cellcolor[HTML]{C0C0C0}0.8445 \\ \bottomrule
\end{tabular}
}
\end{minipage}
\begin{minipage}[b]{.31\linewidth}
\renewcommand\arraystretch{1.06}
\resizebox{\textwidth}{!}{%
\centering
\begin{tabular}{@{}ccc@{}}
\multicolumn{3}{c}{\textbf{(b) Network selection.}}                           \\ \midrule
\multicolumn{1}{c|}{\begin{tabular}[c]{@{}c@{}}Modules\\ ablation\end{tabular}} & \multicolumn{1}{c|}{\textbf{PSNR/$dB $}} & \textbf{SSIM$\uparrow $} \\ \midrule
Direct. SR   & 28.05                         & 0.9038                         \\
Denoising+SR & \cellcolor[HTML]{C0C0C0}28.09 & \cellcolor[HTML]{C0C0C0}0.9072 \\ \bottomrule
\end{tabular}
}
\end{minipage}
\begin{minipage}[b]{.31\linewidth}
\renewcommand\arraystretch{0.9}
\resizebox{\textwidth}{!}{%
\centering
\begin{tabular}{@{}ccc@{}}
\multicolumn{3}{c}{\textbf{(c) RCA Block.}}                         \\ \midrule
\multicolumn{1}{c|}{\begin{tabular}[c]{@{}c@{}}Denoising\\ head\end{tabular}} & \multicolumn{1}{c|}{\textbf{PSNR/$dB $}} & \textbf{SSIM$\uparrow $} \\ \midrule
DNCNN & 27.34                         & 0.8747                         \\
+Att. & \cellcolor[HTML]{C0C0C0}28.78 & \cellcolor[HTML]{C0C0C0}0.9250 \\ \bottomrule
\end{tabular}
}
\end{minipage}
\begin{adjustbox}{center}
\begin{minipage}[b]{0.8\textwidth}
\vspace{0.3em}
\resizebox{\textwidth}{!}{%
\begin{tabular}{@{}cccccc@{}}
\multicolumn{6}{c}{\textbf{(d) Hyperparameters.}}                                             \\ \midrule
\multicolumn{1}{c|}{Metrics} &
  \multicolumn{1}{c|}{\begin{tabular}[c]{@{}c@{}}+dropout\\ (P=0.5)\end{tabular}} &
  \multicolumn{1}{c|}{\begin{tabular}[c]{@{}c@{}}+dropout\\ (P=0.1)\end{tabular}} &
  \multicolumn{1}{c|}{\begin{tabular}[c]{@{}c@{}}+denoising pretrain\\ $\alpha$ = 1e-5/$\beta$=0\end{tabular}} &
  \multicolumn{1}{c|}{\begin{tabular}[c]{@{}c@{}}+SR pretrain\\ $\beta$ = 1e-5/$\alpha$=0\end{tabular}} &
  \begin{tabular}[c]{@{}c@{}}+joint train\\ $\beta$ = 1e-5/$\alpha$=1e-5\end{tabular} \\ \midrule
\textbf{PSNR/$dB $}      & 29.02  & 29.34  & 29.13  & 29.90  & \cellcolor[HTML]{C0C0C0}30.14  \\
\textbf{SSIM$\uparrow $} & 0.9324 & 0.9179 & 0.9236 & 0.9281 & \cellcolor[HTML]{C0C0C0}0.9271 \\ \bottomrule
\end{tabular}

}
\end{minipage}
\end{adjustbox}
\end{table}

\section{Conclusion\label{sec4}}
\wqz{In this study, we presented the AID-SRGAN model for musculoskeletal radiograph super-resolution. In addition, we \hys{introduced} residual channel attention (RCA) block for complex degradation factors. To train the proposed model and adapt to the \hyss{open world} SR task, we further proposed a medical degradation model that included most possible degradation factors, such as Gaussian blur, motion blur, and compression. Finally, \hyss{the experimental results show that the proposed model outperforms its counterparts in terms of PSNR and SSIM.} \hys{In the future, further studies will be carried out to validate degradation models for other medical images.}}

\bibliographystyle{splncs04.bst}
\bibliography{ref}

\begin{thebibliography}{10}
\providecommand{\url}[1]{\texttt{#1}}
\providecommand{\urlprefix}{URL }
\providecommand{\doi}[1]{https://doi.org/#1}

\bibitem{honarvar2021motion}
Asli, H.S., et~al.: Motion blur invariant for estimating motion parameters of
  medical ultrasound images. Scientific Reports  \textbf{11}(1),  1--13 (2021)

\bibitem{Moran2021SR}
Beatriz, M., et~al.: Using super-resolution generative adversarial network
  models and transfer learning to obtain high resolution digital periapical
  radiographs. Computers in biology and medicine  \textbf{129},  104139 (2021)

\bibitem{Chen2022RealWorldSI}
Chen, H., et~al.: Real-world single image super-resolution: A brief review.
  Inf. Fusion  \textbf{79},  124--145 (2022)

\bibitem{bone2021}
Christ, A.B., et~al.: Compliant compression reconstruction of the proximal
  femur is durable despite minimal bone formation in the compression segment.
  Clinical Orthopaedics and Related Research{\textregistered}  \textbf{479}(7),
   1577--1585 (2021)

\bibitem{devlin2018bert}
Devlin, J., Chang, M.W., Lee, K., Toutanova, K.: Bert: Pre-training of deep
  bidirectional transformers for language understanding. arXiv preprint
  arXiv:1810.04805  (2018)

\bibitem{Dimililer2022DCTbasedMI}
Dimililer, K.: Dct-based medical image compression using machine learning.
  Signal, Image and Video Processing  \textbf{16},  55--62 (2022)

\bibitem{de2021impact}
de~Farias, E.C., Di~Noia, C., Han, C., Sala, E., Castelli, M., Rundo, L.:
  Impact of gan-based lesion-focused medical image super-resolution on the
  robustness of radiomic features. Scientific reports  \textbf{11}(1),  1--12
  (2021)

\bibitem{musculoskeletal2020}
Groot, O.Q., et~al.: Does artificial intelligence outperform natural
  intelligence in interpreting musculoskeletal radiological studies? a
  systematic review. Clinical orthopaedics and related research
  \textbf{478}(12), ~2751 (2020)

\bibitem{GuanLCWLZ21}
Guan, M., et~al.: Perceptual quality assessment of chest radiograph. In:
  Medical Image Computing and Computer Assisted Intervention - {MICCAI} 2021
  Proceedings, Part {VII}. vol. 12907, pp. 315--324. Springer (2021).
  \doi{10.1007/978-3-030-87234-2\_30}

\bibitem{Heller2021EducationalCA}
Heller, T., et~al.: Educational content and acceptability of training using
  mobile instant messaging in large hiv clinics in malawi. Annals of Global
  Health  \textbf{87} (2021). \doi{10.5334/aogh.3208}

\bibitem{Hu2020SE}
Hu, J., et~al.: Squeeze-and-excitation networks. IEEE Transactions on Pattern
  Analysis and Machine Intelligence  \textbf{42},  2011--2023 (2020)

\bibitem{huang2021infrared}
Huang, Y., et~al.: Infrared image super-resolution via transfer learning and
  psrgan. IEEE Signal Processing Letters  \textbf{28},  982--986 (2021)

\bibitem{kingma2014adam}
Kingma, D.P., Ba, J.: Adam: A method for stochastic optimization. arXiv
  preprint arXiv:1412.6980  (2014)

\bibitem{Lee2022LearningMP}
Lee, S., Ahn, S., Yoon, K.: Learning multiple probabilistic degradation
  generators for unsupervised real world image super resolution. ArXiv
  \textbf{abs/2201.10747} (2022)

\bibitem{liu2018non}
Liu, D., et~al.: Non-local recurrent network for image restoration. Advances in
  neural information processing systems  \textbf{31} (2018)

\bibitem{ma2021structure}
Ma, C., et~al.: Structure-preserving image super-resolution. IEEE Transactions
  on Pattern Analysis and Machine Intelligence pp.~1--1 (2021).
  \doi{10.1109/TPAMI.2021.3114428}

\bibitem{Mohan2020SABERAS}
Mohan, K.A., Panas, R.M., Cuadra, J.A.: Saber: A systems approach to blur
  estimation and reduction in x-ray imaging. IEEE Transactions on Image
  Processing  \textbf{29},  7751--7764 (2020)

\bibitem{miccaiPengZC21}
Peng, C., Zhou, S.K., Chellappa, R.: {DA-VSR:} domain adaptable volumetric
  super-resolution for medical images. In: Medical Image Computing and Computer
  Assisted Intervention - {MICCAI} 2021. Lecture Notes in Computer Science,
  vol. 12906, pp. 75--85. Springer (2021),
  \url{https://doi.org/10.1007/978-3-030-87231-1\_8}

\bibitem{PengYLY20}
Peng, H., et~al.: Secure and traceable image transmission scheme based on
  semitensor product compressed sensing in telemedicine system. {IEEE} Internet
  Things J.  \textbf{7}(3),  2432--2451 (2020)

\bibitem{rajpurkar2017mura}
Rajpurkar, P., et~al.: Mura: Large dataset for abnormality detection in
  musculoskeletal radiographs. arXiv preprint arXiv:1712.06957  (2017)

\bibitem{Rezaei2017DeepLF}
Rezaei, M., Yang, H., Meinel, C.: Deep learning for medical image analysis.
  ArXiv  \textbf{abs/1708.08987} (2017)

\bibitem{von2021multitask}
von Schacky, C.E., et~al.: Multitask deep learning for segmentation and
  classification of primary bone tumors on radiographs. Radiology
  \textbf{301}(2),  398--406 (2021)

\bibitem{Selvaraju2019GradCAMVE}
Selvaraju, R.R., et~al.: Grad-cam: Visual explanations from deep networks via
  gradient-based localization. International Journal of Computer Vision
  \textbf{128},  336--359 (2019)

\bibitem{simonyan2014very}
Simonyan, et~al.: Very deep convolutional networks for large-scale image
  recognition. arXiv preprint arXiv:1409.1556  (2014)

\bibitem{van2020super}
van Sloun, R.J., Solomon, O., Bruce, M., Khaing, Z.Z., Wijkstra, H., Eldar,
  Y.C., Mischi, M.: Super-resolution ultrasound localization microscopy through
  deep learning. IEEE transactions on medical imaging  \textbf{40}(3),
  829--839 (2020)

\bibitem{Son2021TowardRS}
Son, S., et~al.: Toward real-world super-resolution via adaptive downsampling
  models. IEEE Transactions on Pattern Analysis and Machine Intelligence
  (2021). \doi{10.1109/TPAMI.2021.3106790}

\bibitem{CT2019review}
Thanh, D.N.H., et~al.: A review on {CT} and x-ray images denoising methods.
  Informatica (Slovenia)  \textbf{43}(2) (2019). \doi{10.31449/inf.v43i2.2179}

\bibitem{Wang2021UnsupervisedDR}
Wang, L., et~al.: Unsupervised degradation representation learning for blind
  super-resolution. 2021 IEEE/CVF Conference on Computer Vision and Pattern
  Recognition (CVPR) pp. 10576--10585 (2021)

\bibitem{Wang2018ESRGANES}
Wang, X., et~al.: {ESRGAN:} enhanced super-resolution generative adversarial
  networks. In: Computer Vision - {ECCV} 2018 Workshops Proceedings, Part {V}.
  Lecture Notes in Computer Science, vol. 11133, pp. 63--79. Springer (2018).
  \doi{10.1007/978-3-030-11021-5\_5}

\bibitem{wang2021real}
Wang, X., et~al.: Real-esrgan: Training real-world blind super-resolution with
  pure synthetic data. In: Proceedings of the IEEE/CVF International Conference
  on Computer Vision. pp. 1905--1914 (2021)

\bibitem{motion2020patient}
Yeung, A., et~al.: Patient motion image artifacts can be minimized and
  re-exposure avoided by selective removal of a sequence of basis images from
  cone beam computed tomography data sets: a case series. Oral Surgery, Oral
  Medicine, Oral Pathology and Oral Radiology  \textbf{129}(2),  e212--e223
  (2020)

\bibitem{Zhang2017dncnn}
Zhang, K., et~al.: Beyond a gaussian denoiser: Residual learning of deep cnn
  for image denoising. IEEE Transactions on Image Processing  \textbf{26},
  3142--3155 (2017)

\bibitem{Zhang2018LearningAS}
Zhang, K., et~al.: Learning a single convolutional super-resolution network for
  multiple degradations. 2018 IEEE/CVF Conference on Computer Vision and
  Pattern Recognition pp. 3262--3271 (2018)

\bibitem{Zhang2019DeepPS}
Zhang, K., et~al.: Deep plug-and-play super-resolution for arbitrary blur
  kernels. 2019 IEEE/CVF Conference on Computer Vision and Pattern Recognition
  (CVPR) pp. 1671--1681 (2019)

\end{thebibliography}
\end{document}